\documentclass[preprint,12pt]{elsarticle}

\usepackage{amssymb}
\usepackage{multicol}
\usepackage{graphicx}
\usepackage{booktabs}
\usepackage{bm,mathrsfs,bbm,amscd}
\usepackage[tbtags]{amsmath}
\usepackage{lastpage}
\usepackage{verbatim}

\begin{document}

\begin{frontmatter}



\title{Uncertainties analysis of fission fraction for reactor antineutrino experiments\tnoteref{label1}}
 \tnotetext[label1]{Corresponding author}
\author{X.B.Ma\corref{cor1}\fnref{label2}}
 \ead{maxb@ncepu.edu.cn}
\author[label2]{F.Lu}
\author[label2]{L.Z.Wang}
\author[label2]{Y.X.Chen}
\author[label3]{W.L.Zhong}
\author[label3]{F.P.An}
\address[label2]{North China Electric Power University, Beijing,102206,China \fnref{label2}}
\address[label3]{Institute of High Energy Physics, Chinese Academy of Sciences, Beijing 100049, China \fnref{label3}}
\begin{abstract}
Reactor antineutrino experiment are used to study neutrino oscillation, search for signatures of nonstandard neutrino interaction, and monitor reactor operation for safeguard application. Reactor simulation is an important source of uncertainties for a reactor neutrino experiment. Commercial code is used for reactor simulation to evaluate fission fraction in Daya Bay neutrino experiment, but the source code doesn't open to our researcher results from commercial secret. In this study, open source code DRAGON was improved to calculate the fission rates of the four most important isotopes in fissions, $^{235}$U,$^{238}$U,$^{239}$Pu and $^{241}$Pu, and then was validated for PWRs using the Takahama-3 benchmark. The fission fraction results are consistent with those of MIT's results. Then, fission fraction of Daya Bay reactor core was calculated by using improved DRAGON code, and the fission fraction calculated by DRAGON agreed well with these calculated by SCIENCE. The average deviation less than 5\% for all the four isotopes. The correlation coefficient matrix between $^{235}$U,$^{238}$U,$^{239}$Pu and $^{241}$Pu were also studied using DRAGON, and then the uncertainty of the antineutrino flux by the fission fraction was calculated by using the correlation coefficient matrix. The uncertainty of the antineutrino flux by the fission fraction simulation is 0.6\% per core for Daya Bay antineutrino experiment. The uncertainties source of fission fraction calculation need further to be studied in the future.

\end{abstract}


\end{frontmatter}

\section{Introduction}
Reactor antineutrino experiment have always acted as an important role in the subject of neutrino physics and relative physics. For instant,  the Savannah River Experiment\cite{lab1,lab1-1} by Reines and Cowan in 1956 first detected the neutrino. The KamLAND\cite{lab3} experiment confirmed neutrino oscillation and explained the solar neutrino deficit together with the SNO experiment in the first few years after 2000. Just before that, the CHOOZ \cite{lab2} experiment determined the most stringent upper limit of the last unknown neutrino mixing angle, $sin^{2}2\theta_{13}<0.17$ at 90$\%$ confidence level. After this, a generation of reactor neutrino experiments\cite{lab4,reno,chooze} made efforts to determine the value of $\theta_{13}$. In March of 2012, the Daya Bay collaboration\cite{lab4} discovered a non-zero value for $sin^{2}2\theta_{13}$ at a 5$\sigma$ confidence level, which has fueled discussions about the direction of neutrino physics in the foreseeable future. At the same time,
The Nucifer Experiment\cite{nucifer} is a proposed test of equipment and methodologies for using neutrino detection (or, more specifically, antineutrino detection) for the monitoring of nuclear reactor activity and the assessment of the isotopic composition of reactor fuels for non-proliferation treaty compliance monitoring. In the future,JUNO\cite{juno} will be determining neutrino mass hierarchy by precisely measuring the energy spectrum of reactor electron antineutrinos.

The prediction of antineutrino flux and its uncertainty is an indispensable part of reactor neutrino experiments, especially absolute measurement experiments which use a single detector. Usually, the following formula is used to calculate the antineutrino flux from one reactor core:
\begin{equation}\label{eq1}
S(E_{\nu})=\frac{W_{th}}{\sum_{i}(f_{i}/F)E_{i}}\sum\limits_{i}(f_{i}/F)S_{i}(E_{\nu})
\end{equation}
Where  $W_{th}$(MeV/s) is the thermal power of the core, $E_{i} $(MeV/fission)\cite{maxb} is the energy released per fission for isotope $i$, $f_{i}$ is the fission rate of isotope $i$, and $F$ is the sum of  $f_{i}$ for all isotopes. Thus, $f_{i}/F$ is the fission fraction of each isotope. $S_{i}(E_{\nu})$ is the antineutrino energy spectrum of isotope $i$, which is normalized to one fission.
Normally, $W_{th}$ and $f_{i}/F$ of each isotope are supplied by the nuclear power plants of the reactor neutrino experiments. $f_{i}/F$ of each isotope are calculated by using commercial reactor simulation code SCIENCE in Daya Bay antineutrino experiment. To evaluate the uncertainties of the fission fraction, DRAGON code was improved to have the ability output the fission fraction and Takahama-3 burnup benchmark was used to verify the correct of the development. Then, the improved DRAGON code is also used and the results calculated by DRAGON are compared with those of SCIENCE.

\section{Improved DRAGON code}
Oscillation experiments detect antineutrinos via the signal:
\begin{equation}\label{eq1}
\bar{\nu_{e}}+p\rightarrow e^{+} + n,
\end{equation}
which has a threshold at 1.8MeV. Reactors produce antineutrinos above this threshold through the decay chains of four primary fissile nuclide: $^{235}$U, $^{238}$U, $^{239}$Pu and $^{141}$Pu. However, DRAGON \cite{dragon} is capable of simulating all the important actinides and some of the important fission products produced during the evolution of a reactor core. These include, but are not limited to, the long-lived isotopes: $^{238}$Pu, $^{240}$Pu, $^{242}$Pu, $^{237}$Np, $^{241}$Am, $^{242}$Cm et.al. DRAGON is a lattice code which means it can be used for cell or assembly transport calculation and depletion calculation, but it can't be used for core calculation. If one want to do the core calculation, DONJON\cite{donjon} must be used and the cross section used in DONJON is generated by DRAGON.
DRAGON is an open-source simulation package that allows one to study the behavior of neutrons in a nuclear reactor. It allows one to determine the isotopic concentrations of radionuclides during the burn-up cycle, as well as to perform isotopic depletions. But, DRAGON can not be used for reactor antineutrino experiment uncertainty analysis directly because it can't calculate fission fraction of $^{235}$U,$^{238}$U,$^{239}$Pu and $^{141}$Pu which are needed in the antineutrino reactor experiment. In order to evaluate the fission fraction for the antineutrino experiment, DRAGON was improved to calculate the fission rates of the four most important isotopes in fissions and two step calculation method was used to the depletion calculation. After that, the correlation coefficients between $^{235}$U, $^{238}$U, $^{239}$Pu and $^{241}$Pu was studied using DRAGON for the first time, and the results was used by the Daya Bay Collaboration the uncertainties analysis of the fission fraction.

\section{The Takahama-3 burnup benchmark}

\begin{figure}
\begin{center}
\includegraphics[width=7cm]{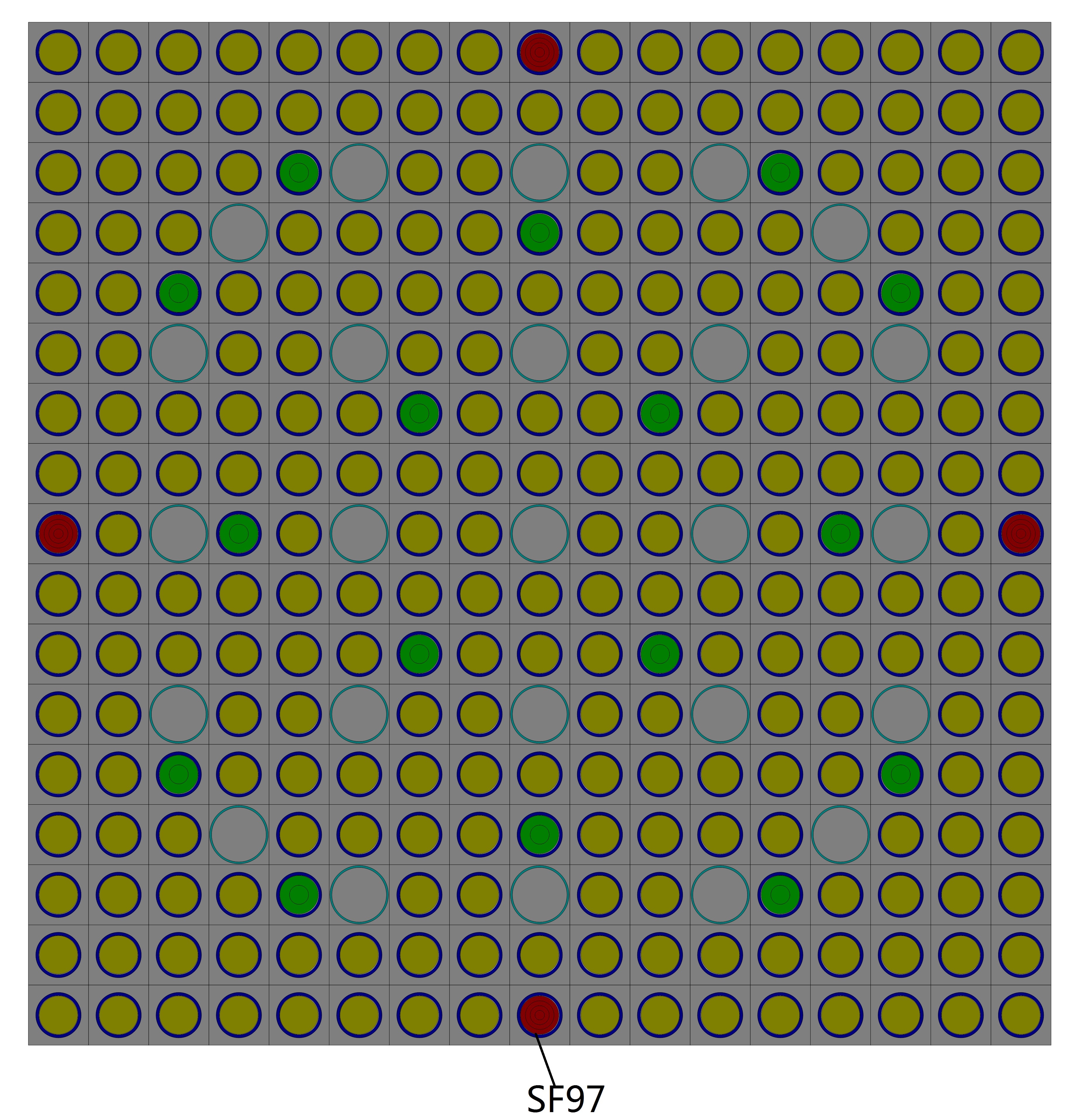}
\caption{Positions of assay fuel rods in Takahama-3 SF97 assemblies}
\label{fig1}
\end{center}
\end{figure}

In order to verify the accuracy of results calculated by DRAGON, the Takahama-3 burnup benchmark was calculated. The Takahama-3 reactor is a PWR as well as Daya Bay reactor, which operates with 157 fuel assemblies producing a total thermal power of 2652MW. The assembly is a 17$\times$17 design, meaning there are 289 locations for rods. A diagram of a Takahama-3 assembly is shown in Fig.\ref{fig1}. The benchmark began with assemblies loaded with fresh UO$_{2}$ fuel rods with an initial enrichment of 4.11\% $^{235}$U by weight, with the remainder being $^{238}$U. Each assembly features 14 gadolinium-bearing(Gd$_{2}$O$_{3}$) fuel rods containing 2.63\% $^{235}$U and 6\% gadolinium by weight.
Sixteen samples selected from three fuel rods irradiated in assemblies NT3G23 and NT3G24 of the Takahama-3 reactor were included for destructive isotopic analysis. The burnup of these samples was between 14 and 47 GWd/MTU.
 Isotopic dilution mass spectroscopy was used to determine uranium and plutonium inventories, different mass spectroscopy and alpha and gamma counting techniques were used to determine isotopic concentrations of the other elements. For the most relevant isotopes, namely $^{235}$U, $^{238}$U, $^{239}$Pu and $^{141}$Pu, the uncertainty associated with the determination of the isotopic mass fraction is $<$0.1\% for uranium isotopes and $<$0.3\% for plutonium isotopes\cite{lab7}.The three fuel rods came from two different assemblies. From the first assembly, labeled NT3G23, a normal uranium dioxide fuel rod (SF95) and a gadolinium-bearing fuel rod (SF96) were studied after two cycles. From the second assembly, labeled NT3G24, a normal uranium dioxide fuel rod (SF97) was studies after three cycles.
We focus on SF97, because it had the longest irradiation time and therefore any cumulative systematic effects would be maximized. The rod was present in three consecutive fuel cycles of 385, 402, and 406 days with 88 days and 62 days of cool-down time between cycles as shown in Table 1. Samples were taken from SF97 at the six locations indicated in Table 1. Sample SF97-1 was located only 163 mm from the top of the rod, making the correct modeling of neutron leakage difficult.
The construction of the SF97 rod simulation starts with a geometric description of the fuel assembly and the initial isotopic inventory of the fuel pellets. The primary inputs used in the simulation are found in Table1 and Table 2, and a mean boron concentration of 630 ppm per cycles\cite{lab8} is used. The pellet stack density is 10.07 g/cm$^{3}$. Those are the standard values used by the other simulations considered. Cross section libraries based on ENDF/B-VII.0 are used in the simulation.

\begin{table}
\begin{center}
\caption{ \label{tab1}  Operation history of Takahama-3}
\footnotesize
\begin{tabular*}{80mm}{c@{\extracolsep{\fill}}cccc}
\toprule Start & Stop   & Days  & Status & Cycle \\
\hline
1990/1/26 &	1991/2/15 &	385  &  Burnup &5 \\
1991/2/15 &	1991/5/14 &	88 	 &   Cool  &   \\
1991/5/14 &	1992/6/19 &	402  &  Burnup &6  \\
1992/6/19 &	1992/8/20 &	62   &   Cool  &   \\
1992/8/20 &	1993/9/30 &	406  &  Burnup &   7 \\
\bottomrule
\end{tabular*}
\end{center}
\end{table}

\begin{table}
\begin{center}
\caption{ \label{tab2} Design data for Takahama-3 reactor and fuel assemblies}
\footnotesize
\begin{tabular*}{80mm}{c@{\extracolsep{\fill}}cc}
\toprule Parameter & Value \\
\hline
Moderator Density &	0.72 g/cm$^3$ \\
Moderator Temperature &	600.0 K \\
Cladding  Temperature &	600.0 K \\
Fuel Temperature & 900.0 K \\
Fuel Density  & 10.07 g/cm$^3$ \\
Fuel Cell Mesh & 1.259 cm \\
Fuel Rod Radius  & 0.4025 cm \\
Fuel Cladding Radius & 0.475 cm \\
Guide Tube Inner Radius & 0.573 cm \\
Guide Tube Outer Radius & 0.613 cm \\

\bottomrule
\end{tabular*}
\end{center}
\end{table}

\begin{figure}
\begin{center}
\includegraphics[width=9cm]{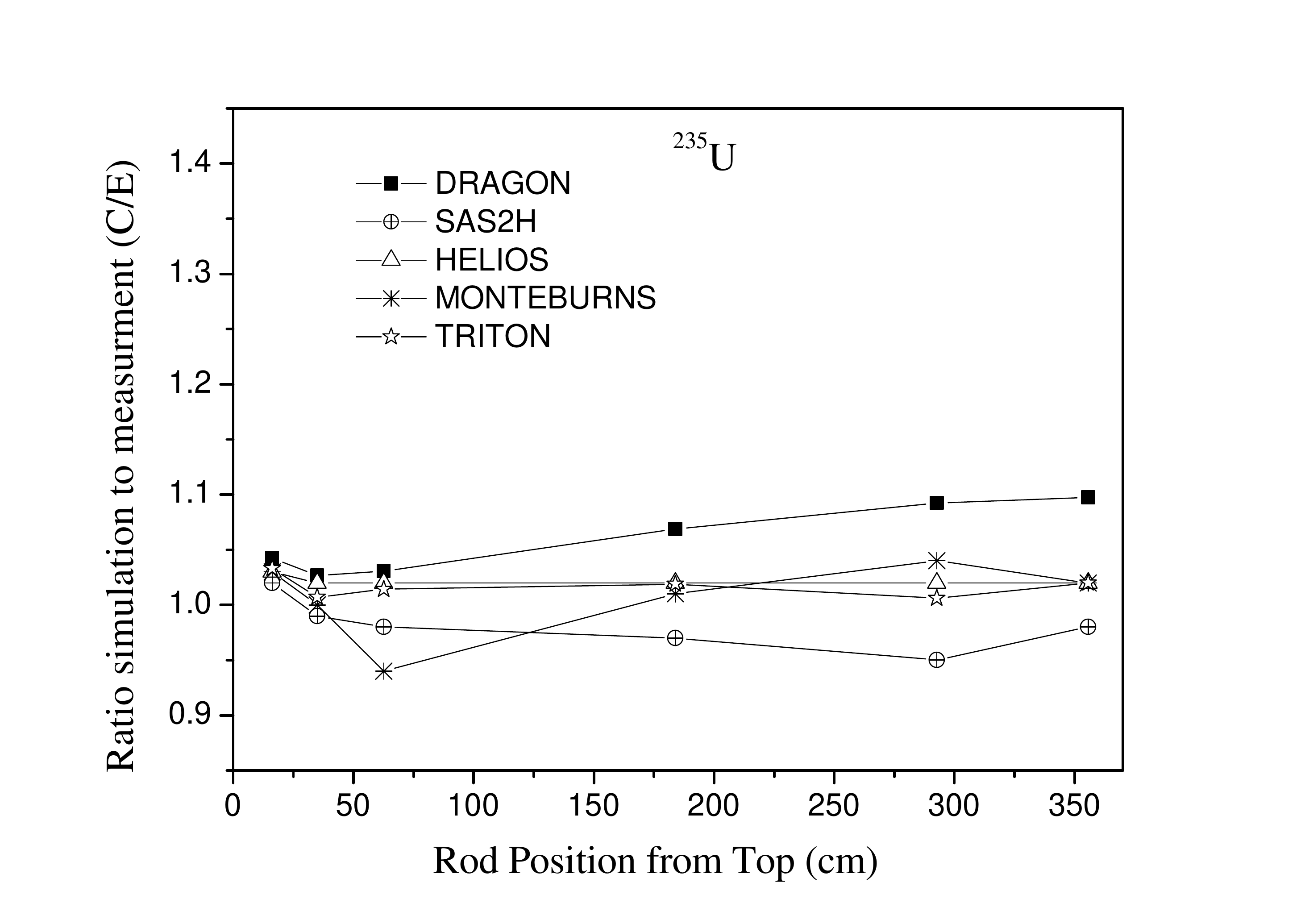}
\figcaption{\label{U235}Comparison of the ratio of calculated to measured mass inventories for SF97 for $^{235}$U  }
\end{center}
\end{figure}

\begin{figure}
\begin{center}
\includegraphics[width=9cm]{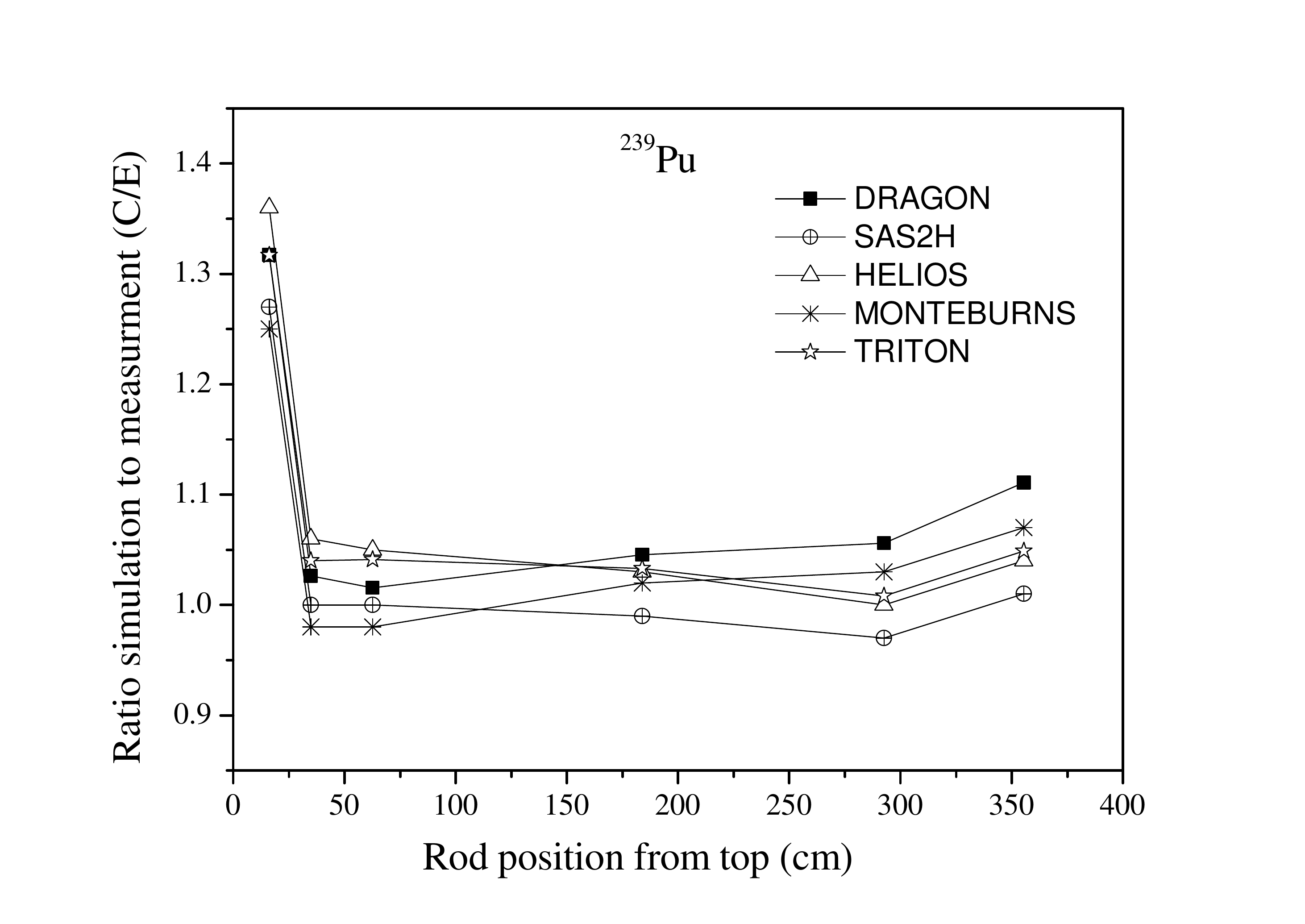}
\caption{\label{Pu239}Comparison of the ratio of calculated to measured mass inventories for SF97 for $^{239}$Pu  }
\end{center}
\end{figure}

\begin{figure}
\begin{center}
\includegraphics[width=9cm]{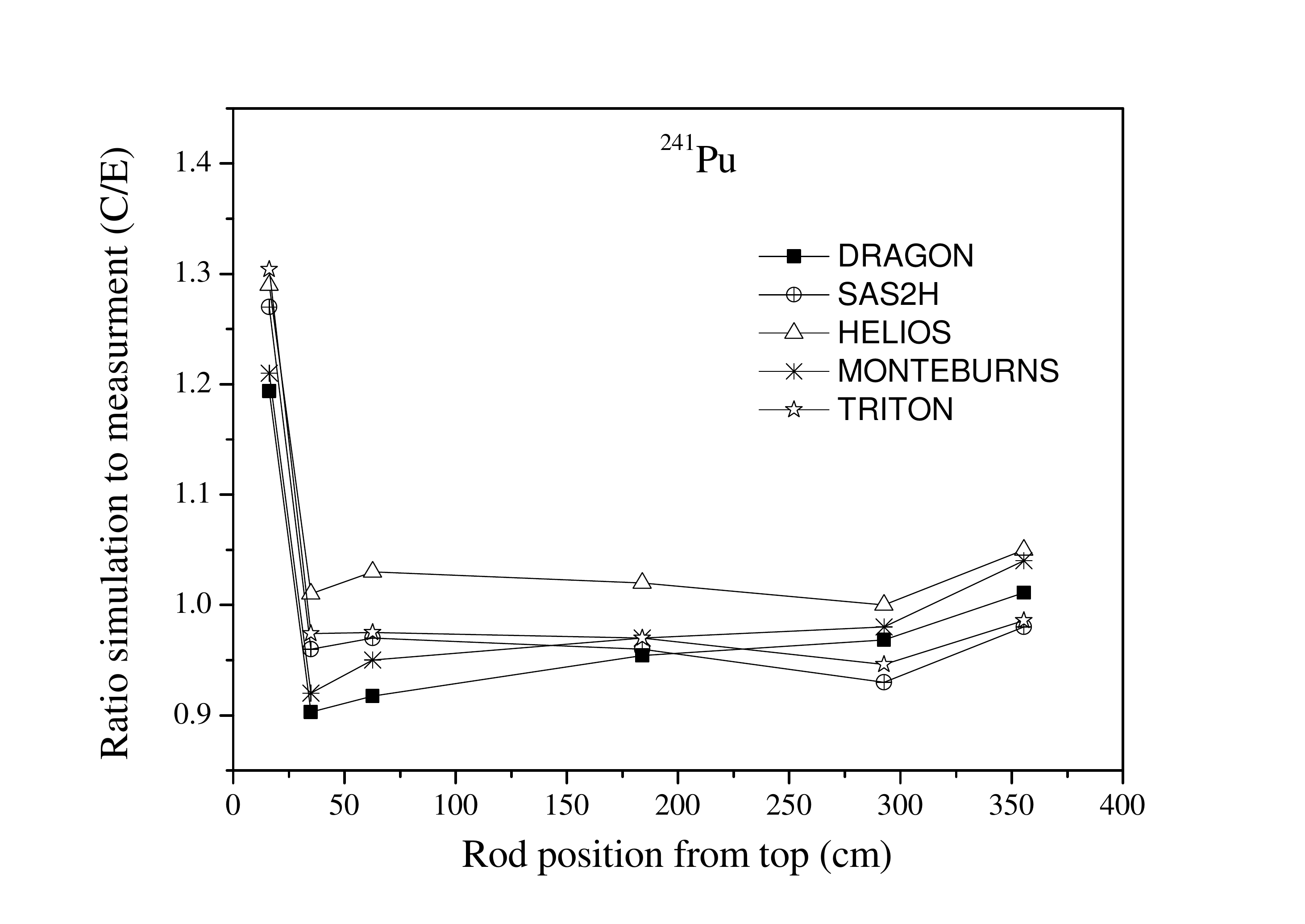}
\caption{\label{Pu241}Comparison of the ratio of calculated to measured mass inventories for SF97 for $^{241}$Pu  }
\end{center}
\end{figure}

\begin{figure}
\begin{center}
\includegraphics[width=9cm]{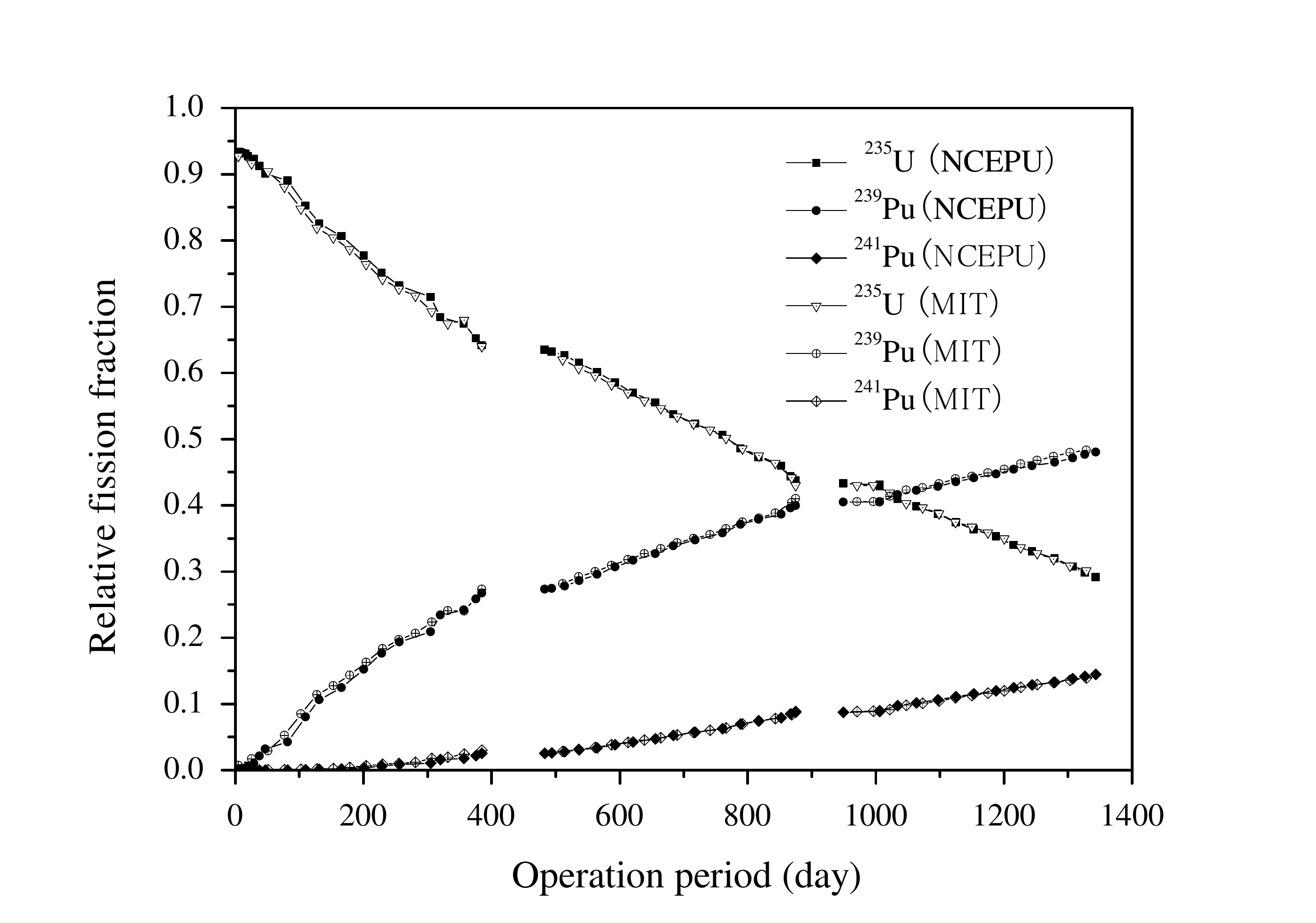}
\caption{\label{fissionrate}Fission fraction from DRAGON for the Takahama simulation. (NCEPU represents our calculation results and MIT represents results of Massachusetts Institute of Technology)}
\end{center}
\end{figure}

When the DRAGON simulation is complete, the results for rod SF97 are extracted. Fig.\ref{U235},\ref{Pu239} and \ref{Pu241} show the ratio of calculated to experimentally measured mass inventories for SF97 for three isotopes important to antineutrino experiments:$^{235}$U, $^{239}$Pu and $^{241}$Pu. For the other isotopes, the DRAGON results are consistent with those calculated by other codes, such as SCALE4.4a\cite{lab8}, SCALE5\cite{lab10}, ORIGEN-S\cite{lab11}, MONTE-BURNS\cite{lab12}, and HELIOS\cite{lab8}.
However, there is a large deviation in SF97-1, located near the top edge of the fuel rod characterized by high leakage and large flux gradients. This effect is observed in results from all the codes which are all based on two dimensional calculation. If one want to take into account of the leakage, three dimensional calculation must be done.
Neglecting SF97-1, we calculated the average deviation over the rod by taking the average of the samples. Even neglecting sample 1, the average deviation for $^{235}$U, $^{239}$Pu and $^{241}$Pu which are shown in Table \ref{tab3} are 6.32\%, 5.08\% and 4.92\% respectively.
Neutrino experiments are interested in fission fraction rather than mass inventories. The instantaneous fission fraction predictions for SF97 through the three fuel cycles was shown in Fig.\ref{fissionrate}. The trend is same as Ref\cite{lab7}. This demonstrate that DRAGON can offer reliable fission fraction for PWRs simulation.

\begin{table}
\begin{center}
\caption{ \label{tab3}  The ratio of calculated to measured mass inventories for SF97.}
\vspace{-3mm}
\footnotesize
\begin{tabular*}{170mm}{@{\extracolsep{\fill}}ccccccccc}
\toprule
Nuclide	& & & & Burnup& (GW.d/tU) & & Considering & Average \\
& 17.69& 30.73&	42.16 & 47.03 & 47.25 &	40.79 &  SF97-1 &  Neglecting SF97-1 \\
\hline
$^{235}$U &1.0424	&1.0265&1.0308&	1.0687&	1.0925&	1.0974&	1.0597&	1.0632   \\
$^{239}$Pu &1.3172 &1.0263&1.0156&1.0455&	1.0558&	1.1108&	1.0952&	1.0508  \\
$^{241}$Pu &1.1939 &0.9030&0.9174&0.9542&	0.9683&	1.0110&	0.9913&	0.9508  \\
\bottomrule
\end{tabular*}%
\end{center}
\end{table}

\section{Antineutrino flux and uncertainties analysis}
\label{sysun}
\begin{figure}
\begin{center}
\includegraphics[width=9cm]{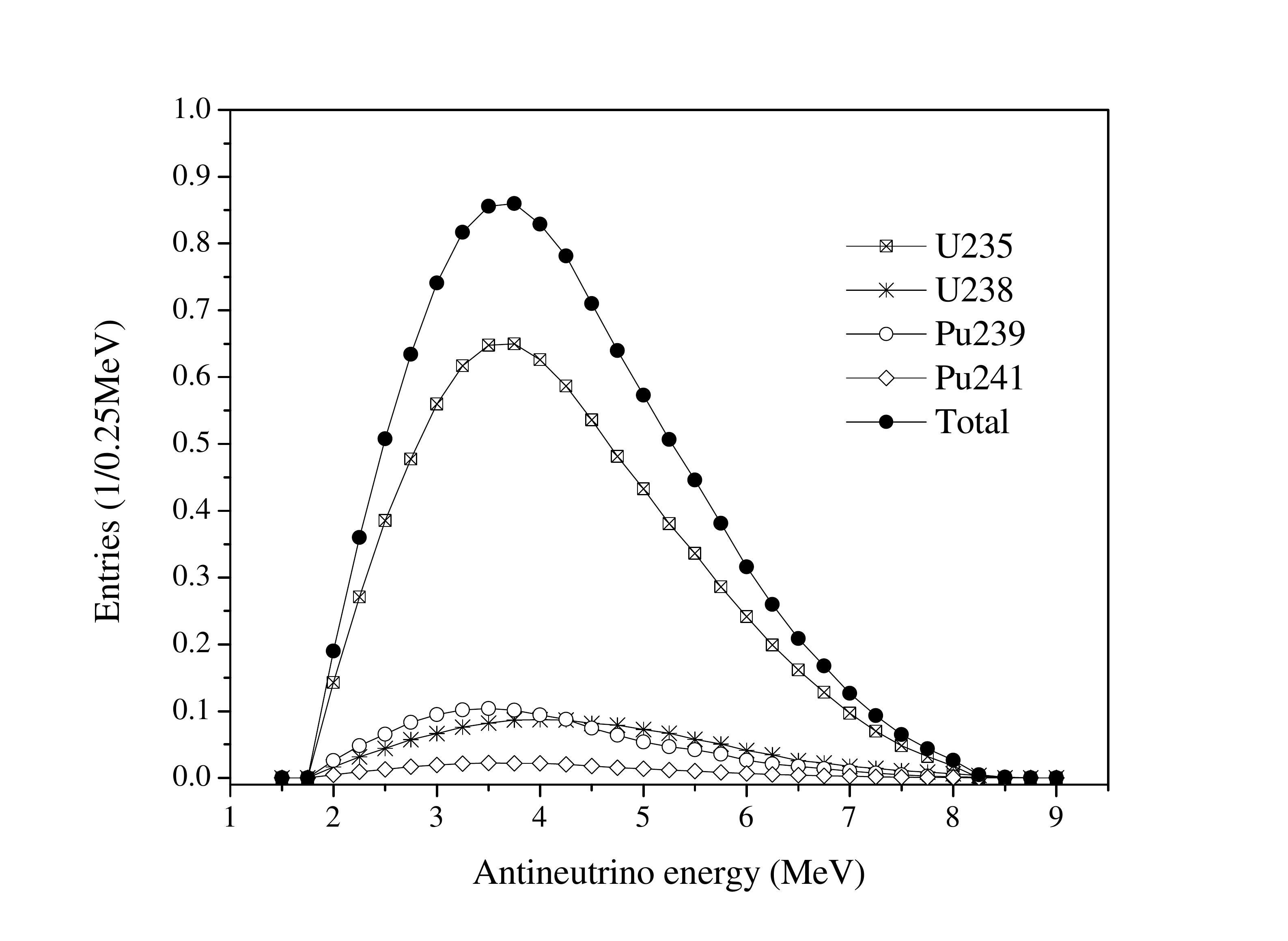}
\caption{\label{spectra}Antineutrino spectra of different isotopes.}
\end{center}
\end{figure}

 If we know the fission rates of each isotope from core simulation, and neutrino energy spectra of each isotope, we can easily get the neutrino flux $S(E_{\nu})=\Sigma_{i}f_{i}S_{i}(E_{\nu})$, where $f_{i}$ is the fission rate of isotope i and $S(E_{\nu})$ is its neutrino spectrum. However, the fission rates are proportional to the thermal power of the core ,which is fluctuating. It is unrealistic to repeat core simulation to reflect the power fluctuation. Normally we scale the neutrino flux to the measured thermal power and Equation(\ref{eq1}) is used to calculate the antineutrino spectra emitted from reactor.
 If the antineutrino spectra which have been emitting from the reactor are known, the antineutrino spectra of the detector can be calculated using the following equation.
\begin{equation}
\label{spemu}
S_{p}(E_{\nu})= \frac{1}{4\pi{L^{2}}}S(E_{\nu})\cdot\sigma_{IBD}(E_{\nu})\cdot{N_{p}}
\end{equation}
Where the detector is located in p position, $S_{p}(E_{\nu})$ are the antineutrino spectra of the detector, L is the length of the baseline, $\sigma_{IBD}(E_{\nu})$ is the cross section of the inverse bate decay reaction, and ${N_{p}}$ is the number of proton in the detector. If the value of different parameter which was given in table \ref{antineu} was supposed as the input parameter when we calculated the antineutrino spectra of the detector, and the fission fraction was obtained by using DRAGON, the antineutrino spectra of the detector was shown in Fig.\ref{spectra}.

In this article, We calculated the fission fraction using DRAGON and compared its value with those from SCIENCE program. If the fission fraction uncertainty was known, then the uncertainty of the antineutrino flux by the fission fraction was obtained using the error transfer formula.
\begin{equation}
\label{uncer}
\frac{\delta R}{R}= \frac{1}{R}\sqrt{\sum\limits_{i,j}\frac{\partial R}{\partial \alpha_{i}}\frac{\partial R}{\partial \alpha_{j}}\cdot\delta\alpha_{i}\delta\alpha_{j}\rho_{i,j}}
\end{equation}
Where $\frac{\delta R}{R}$ is the uncertainty of antineutrino flux by the fission fraction, $\delta\alpha_{i}$ is the uncertainty of fission fraction of the isotopes i (i=$^{235}$U,$^{238}$U,$^{239}$Pu and $^{241}$Pu), $\rho_{i,j}$ are the correlation coefficients between $^{235}$U,$^{238}$U,$^{239}$Pu and $^{241}$Pu.
\begin{table}
\begin{center}
\caption{ \label{antineu}  partly input parameter  }
\footnotesize
\begin{tabular*}{80mm}{c@{\extracolsep{\fill}}cccc}
\toprule
Parameter & $W_{th} (GW)$   & L (Km)  & $N_{p}$  \\
\hline
Value &	2.85 &	1.0  &  7.695$\times 10^{28}$  \\
\bottomrule
\end{tabular*}
\end{center}
\end{table}

\begin{figure}
\begin{center}
\includegraphics[width=9cm]{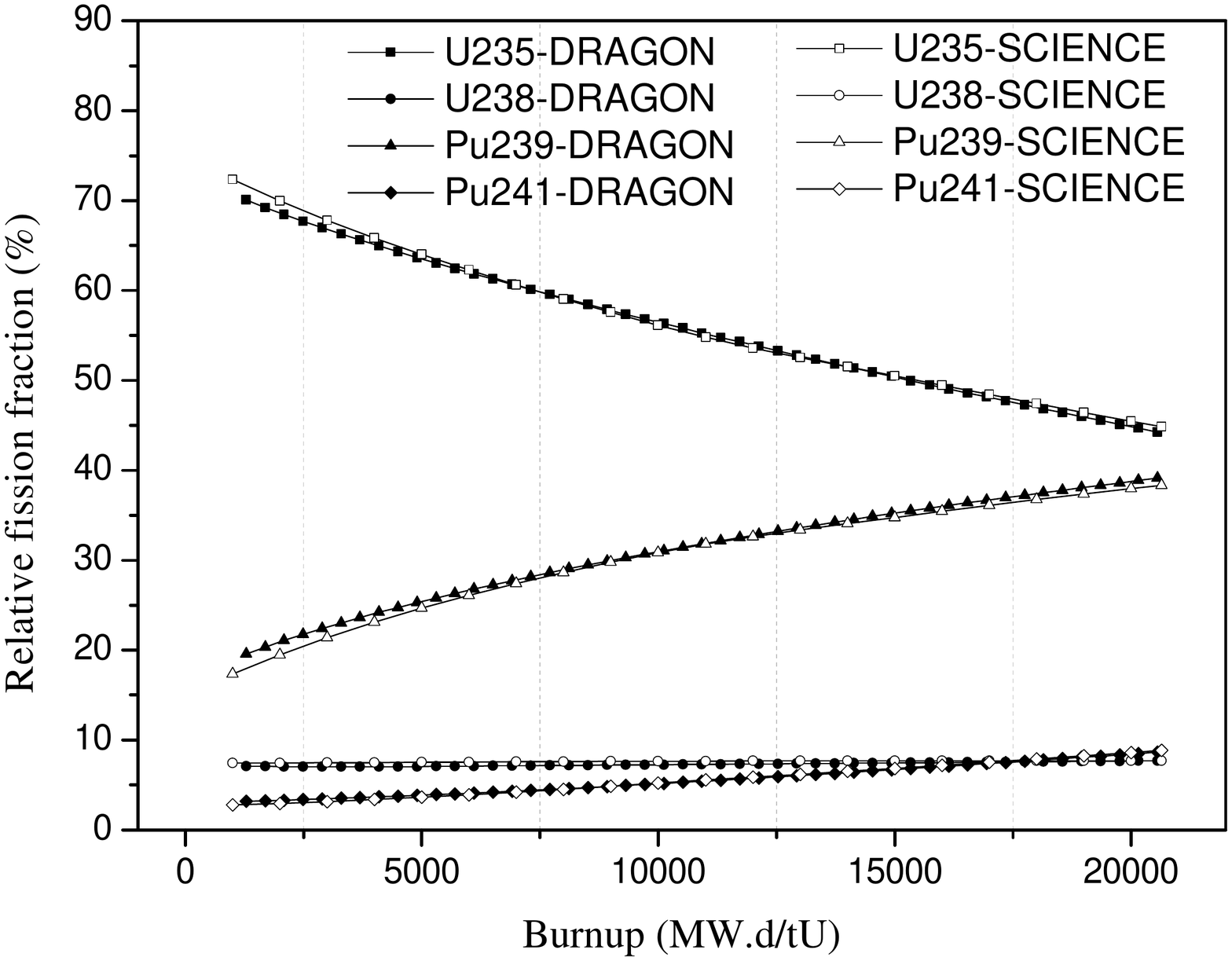}
\caption{\label{fissionduibi}Fission fraction from DRAGON and SCIENCE.}
\end{center}
\end{figure}

The reactor core of the Daya Bay nuclear power plant consists of 157 fuel assemblies, and each assembly contains 264 fuel cell of uranium dioxide. For the initial cores, there were three kinds of enrichment fuel in the core, such as 1.8\%, 2.4\%, and 3.1\%, and the refueling cycle period was 12 moths. For now Daya Bay has increased their fuel enrichment to 4.45\% in order to archive refueling cycle from 12 moths to 18 moths. By the end of the cycle, about 1/3 assemblies will be discharged from the core and the fuel in the outer region will be moved to the inner region to get a deeper buren-up. In this article, DRAGON was used to simulate the core depletion and relative fission fraction from DRAGON was compared with that from SCIENCE, which was used in the Daya Bay nuclear power plant. The results were shown in Fig.\ref{fissionduibi}. For the four isotopes, the DRAGON results were consistent with SCIENCE calculation results. The average deviation for $^{235}$U,$^{238}$U, $^{239}$Pu and $^{241}$Pu are 0.71\%, 4.2\%, 2.1\% and 3.5\% respectively. The average deviation for all isotopes are less than the deviation which was calculated using takahama benchmark, but the results which obtained from the experiment benchmark was usually regarded as more reliable than that obtained from the different program comparing. In addition, in reference \cite{apollo}, the interpretation performed with the APOLLO2.8 code package shows that the concentrations of the main fissile isotopes $-$ $^{235}$U, $^{239}$Pu and $^{241}$Pu $-$ are predicted with an experimental uncertainty below 5\% (1$\sigma$). Therefor, If we assume the uncertainty of the fission fraction is proportional to the uncertainty of the isotopes density, it is suitable that the uncertainty of the fission fraction could be estimated as 5\%.

The correlation coefficients between $^{235}$U,$^{238}$U,$^{239}$Pu and $^{241}$Pu was important for determined the uncertainty of the antineutrinos flux by the fission fraction. The correlation coefficients were evaluated using assembly data from the Daya Bay reactor core. Table \ref{tab1} show the correlation coefficient matrix between $^{235}$U,$^{238}$U,$^{239}$Pu and $^{241}$Pu. According to the difference of the calculation atomic density and measurement by different programs, correlation coefficients are also obtained from reference \cite{zdj}. Because of using the Daya Bay reactor core data, the results of us is maybe more suit for Daya Bay antineutrino experiment.

 According to the equation (\ref{uncer}), If we take the uncertainties for all isotopes as 5\%, the uncertainty of the antineutrino flux by the fission fraction simulation is 0.6\% per core.
\begin{table}
\begin{center}
\caption{ \label{tab1}  Correlation coefficient matrix between $^{235}$U,$^{238}$U,$^{239}$Pu and $^{241}$Pu}
\footnotesize
\begin{tabular*}{80mm}{c@{\extracolsep{\fill}}ccccc}
\toprule
   & $^{235}$U   & $^{238}$U  & $^{239}$Pu & $^{241}$Pu \\
\hline
$^{235}U$  &	1.00 &	-0.22  &  -0.53 & -0.18 \\
$^{238}$U  &	-0.22 &	1.00 	 &   0.18  &  0.26 \\
$^{239}$Pu &	-0.53 &	0.18  &  1.00 & 0.49  \\
$^{241}$Pu &	-0.18 &	0.26   &   0.49  & 1.00  \\
\bottomrule
\end{tabular*}
\end{center}
\end{table}

\section{Conclusion}

In order to predict the fission fraction of the reactor in the antineutrino experiment, DRAGON was improved and Takahama-3 burnup benchmark was used to verify the uncertainties by comparison calculation results with the experiment. Given the power history of assembles of the reactor core, the improved DRAGON code was used to simulate the Daya Bay reactor and the fission fraction calculated by DRAGON was consistent with that of SCIENCE results. The average deviation for $^{235}$U,$^{238}$U, $^{239}$Pu and $^{241}$Pu are 0.71\%, 4.2\%, 2.1\% and 3.5\% respectively. Comparison of the 5\% uncertainties for each isotopes used in Daya Bay antineutrino experiment, the 5\% uncertainties are suitable and conservative. The uncertainty of the antineutrino flux by the fission fraction simulation is 0.6\% per core for Daya Bay antineutrino experiment.
Otherwise, we still don't know what reason account for the 5\% fission fraction uncertainties. In generally, the reaction rate can be defined as
\begin{equation*}\label{rate}
  R^{i}_{f} = \Sigma^{i}_{f} \times \bar{\phi} = N_{i}\bar{\sigma}^{i}_{f}\bar{\phi}
\end{equation*}
where, $R^{i}_{f}$ is the fission rate of isotopes i, $\Sigma^{i}_{f}$ is the average macroscopic fission cross section, $N_{i}$ is the atomic density of isotopes i, $\bar{\sigma}^{i}_{f}$ is the average microscopic fission cross section, and $\bar{\phi}$ is the average neutron flux.
According to equation (\ref{rate}), the fission fraction can be defined as
\begin{equation*}\label{rate}
  f_{i} = R^{i}_{f}/\sum_{i}(R^{i}_{f})
\end{equation*}
If the uncertainties of $N_{i}$, $\bar{\sigma}^{i}_{f}$ and $\bar{\phi}$ are known, we may have more knowledge of source uncertainties of fission fraction. the uncertainties of $N_{i}$, $\bar{\sigma}^{i}_{f}$ and $\bar{\phi}$ can be done using the core simulation code. The uncertainties of each term should be studied in the future.

\section*{Acknowledgements}
The work was supported by National Natural Science Foundation of China (No. 11175201,11390383). I would like to thank research fellow Cao Jun for its extraordinary support.





\end{document}